\documentstyle[12pt,epsf,twoside,fleqn,espcrc1]{article}

\newcommand{\AmS}{{\protect\the\textfont2
A\kern-.1667em\lower.5ex\hbox{M}\kern-.125emS}} 

\def\simlt{\stackrel{<}{{}_\sim}}

\newcommand{\be}{\begin{eqnarray}}
\newcommand{\ee}{\end{eqnarray}}
\newcommand{\ben}{\begin{eqnarray*}}
\newcommand{\een}{\end{eqnarray*}}

\title{Electromagnetic Emission Rates and Spectral Sum Rules
\thanks{Work supported by the Department of 
Energy under grant No. DE-FG-88ER40388}}

\author{James Steele\address{Department of Physics, The Ohio State
 University, Columbus, OH 43210, USA}
 Hidenaga Yamagishi\address{4 Chome 11-16-502, Shimomeguro, Meguro, Tokyo,
Japan. 153.} and Ismail Zahed\address{Department of Physics,
SUNY, Stony Brook, New York 11794, USA}}

\begin{document}

% typeset front matter
\maketitle

\begin{abstract}
The electromagnetic emission rates at SPS energies satisfy
spectral constraints in leading order in the pion and nucleon 
densities. These constraints follow from the strictures of broken
chiral symmetry. We saturate these constraints using available
data, leading to model independent emission rates from a hadronic
gas. With a simple fire-ball scenario, only large nucleon densities
may account for the present CERES data.
\end{abstract}

\vskip 0.5cm

{\bf 1.}
Recent relativistic heavy-ion collisions at CERN have reported an
excess of dileptons over a broad range of lepton invariant mass
\cite{CERES,HELIOS}. A possible excess was also reported
in the direct photon spectrum \cite{WA80}. 
In this talk, we would like to show that under the assumption that
the heavy-ion collision at SPS energies trigger a hadronic gas,
the photon and dilepton emission rates are constrained by
available data in the vacuum, to leading order in the pion and the 
nucleon density. The density expansion is justified if we note that
in the hadronic gas the expansion parameters are 
$\kappa_\pi = n_\pi/2m_{\pi}f_{\pi}^2 \simlt 0.3$ for the pions,
and $\kappa_N = n_N g_A^2/2m_N f_{\pi}^2\simlt 0.3$ for nucleon
densities $n_N\simlt 3\rho_0$ with $\rho_0\sim 0.17$ fm$^{-3}$
the nuclear matter density \cite{US2}.

{\bf 2.}
In a hadronic gas in thermal equilibrium, the rate ${\bf 
R}$ of dileptons produced in an unit four volume follows from the thermal 
expectation value of the electromagnetic current-current correlation 
function \cite{LARRY}. For massless leptons with momenta $p_1, p_2$,
the rate per unit invariant momentum $q =p_1+p_2$ is given by 
\be
\frac {d{\bf R}}{d^4q} = -\frac{\alpha^2}{6\pi^3 q^2}\,\,
\,\,\frac 2{1+e^{q^0/T}}\,\,{\rm Im}{\bf W}^F (q)
\label{1}
\ee
where $\alpha =e^2/4\pi$ is the fine structure constant, and
\ben
{\bf W}^F (q) = i\! \int\!\! d^4x \, e^{iq\cdot x} \,
{\rm Tr} \left(e^{-({\bf H}-\mu\, {\bf N} -\Omega)/T} \,\,T^*{\bf
J}^{\mu} (x){\bf J}_{\mu} (0)\right). 
\label{4}
\een
$e{\bf J}_{\mu}$ is the hadronic part of the electromagnetic current,
${\bf H}$ is the hadronic Hamiltonian, $\mu$ the baryon chemical
potential, $\bf N$ the baryon number operator, $\Omega$ the Gibbs
energy, $T$ the temperature, and the trace is over a complete set of
hadron states.

{\bf 3.}
For temperatures $T\simlt m_{\pi}$ and baryonic densities $n_N\simlt
3\rho_0$ we may expand the trace in (\ref{4}) using pion and nucleon
states. To first order in the density, we have
\be
{\rm Im} \,{\bf W}^F (q) = -3q^2 {\rm Im} \;{\bf \Pi}_V(q^2)
+\frac1{f_\pi^2}\int\! d\pi\, {\bf W}^F_\pi(q,k) 
+ \int\! dN\, {\bf W}^F_N (q,p) 
+ {\cal O}\!\left(\kappa_\pi^2, \kappa_N^2, \kappa_\pi\kappa_N
\right). 
\label{6}
\ee
The phase space factors are 
\be
dN = \frac {d^3p}{(2\pi)^3} \frac 1{2E_p}
\frac 1{e^{(E_p-\mu)/T} +1}\qquad{\rm and}\qquad
d\pi = \frac {d^3k}{(2\pi)^3} \frac {1}{2\omega_k} 
\frac 1{e^{\omega_k/T} -1}\nonumber
\ee
with the nucleon energy $E_p=\sqrt{m_N^2 +p^2}$ and the pion energy 
$\omega_k=\sqrt{m_{\pi}^2 +k^2}$.  The first term in (\ref{6}) is the
transverse part of the isovector correlator $\langle 0|T^* {\bf V}{\bf V} |0
\rangle$ and summarises the results of the resonance gas model. It is
given by the $e^{+}e^{-}$ annihilation data. At low $q^2$ 
it is dominated by the $\rho, \rho', ...$, while at high $q^2$
its tail is dual to the $q\overline{q}$ spectrum. 

%%%%%%%%%%%%%%%%%%%%%%%%%%%%%%%%%%%%%%%%%%%%%%%%%%%%%%%%%%
\begin{figure}
\vspace{-.75in}
\begin{center}
\leavevmode
\epsfysize=4in
\epsffile{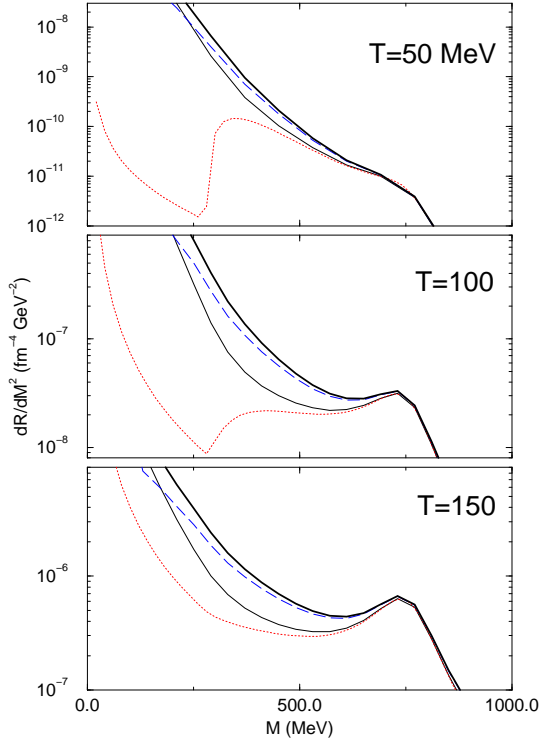}
\end{center}
\caption{\label{dilep.ps}
The dielectron rate for pions alone (dotted), pions and $\Delta$ (solid),
and pions and one-loop (dashed).  The contribution from pions,
$\Delta$, and one-loop together is represented by the thick solid
line.  A fixed nucleon density of $\rho_0$ was used.} 
\end{figure}
%%%%%%%%%%%%%%%%%%%%%%%%%%%%%%%%%%%%%%%%%%%%%%%%%%%%%%%%%%%%%%

The term linear in pion density can be reduced by the use of chiral
reduction formulas to a form amenable to experimental determinations.
The important contributions are \cite{US1}
\be
{\bf W}^F_{\pi}(q,k) \simeq&& 12q^2 {\rm Im}\; {\bf \Pi}_V(q^2)
-6(k+q)^2 {\rm Im}\; {\bf \Pi}_A\left( (k+q)^2 \right) +
(q\rightarrow -q)
\nonumber\\
&&{}+8((k\cdot q)^2-m_\pi^2 q^2) {\rm Im}\;
{\bf \Pi}_V(q^2) 
\times{\rm Re} \left( \Delta_R(k+q) + \Delta_R(k-q) \right)  
\label{7}
\ee
with $\Delta_R (k)$ the retarded pion propagator,
and ${\bf \Pi}_A$ the transverse part of the isoaxial correlator $\langle
0 | T^* {\bf j}_A {\bf j}_A| 0\rangle$ which follows from tau decay
data \cite{US1}.  It is dominated by the $a1$ resonance.

The term linear in the nucleon density is
just the spin-averaged forward Compton scattering amplitude on the
nucleon with virtual photons.  This is only measured for various
values of $q^2 \le 0$.  However, the dilepton and photon rates require
$q^2 \ge 0$.  Therefore, only the photon rate for this term can be
determined directly from data by use of the optical theorem
\be
e^2 {\bf W}^F_N(q,p) =
-4(s-m_N^2) \sum_I \sigma^{\gamma N}_{\rm tot.}(s)
\label{8}
\ee 
with $s=(p+q)^2$.  
For off-shell photons, we must resort to chiral constraints to
determine the nucleon contribution to the dilepton rate.  Broken chiral
symmetry dictates uniquely the form of the strong interaction
Lagrangian (at tree level) for spin $\frac12$. Perturbative unitarity follows
from an on-shell loop-expansion in $1/f_{\pi}$, that enforces 
current conservation and crossing symmetry. To one-loop the contribution
in this case is parameter free. The large contribution of the
$\Delta$ to the Compton amplitude near threshold  is readily 
taken into account by adding it as a unitarized tree term
to the one-loop result \cite{US2}.

Our dilepton rate at $T=150 $MeV is shown in Fig.~\ref{dilep.ps}.
The dominant effect in our case comes from
the continuum and not the $\Delta$ resonance.  At $M=400$ MeV, the
inclusion of nucleons enhances the rate by a factor of three.  Others
who have taken nucleons into account through various methods
\cite{RAPP} find enhancements in the rate similar to our
result. In our calculations there is no shift of the dilepton pair 
production $\rho$ peak.

%%%%%%%%%%%%%%%%%%%%%%%%%%%%%%%%%%%%%%%%%%%%%%%%%%%%%%%%%%
\begin{figure}
\vspace{-1in}
\begin{center}
\leavevmode
\epsfysize=3.75in
\epsffile{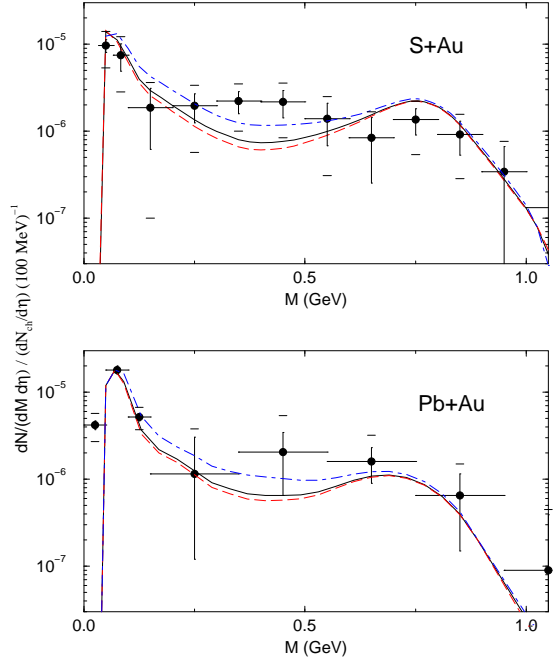}
\end{center}
\vspace{-.35in}
\caption{\label{data.ps} Our dielectron rate including the $\Delta$
and one-loop contributions evolved in space-time as in
\protect\cite{US2} for S-Au 
and Pb-Au collisions.  In the upper graph, $n_N=0,0.7\rho_0$, and
$2.5\rho_0$ are plotted as the dashed, solid, and dashed-dotted lines
respectively.  In the lower graph, the lines are for
$n_N=0,\rho_0$, and $4\rho_0$.  The data are from \protect\cite{CERES}.  The
systematic errors are added linearly to the statistical error bars
to give the cross line.}
\end{figure}
%%%%%%%%%%%%%%%%%%%%%%%%%%%%%%%%%%%%%%%%%%%%%%%%%%%%%%%%%%%%%%

{\bf 4.}
In order to fully understand the role of the experimental cuts, we 
have used a simple fire-ball evolution. The details of these calculations 
are given in \cite{US2}. The results are shown in fig.~\ref{data.ps},
for S-Au and Pb-Au collisions. The Dalitz and prompt 
omega decays were borrowed from the transport model
\cite{BROWN}.  
Adding the nucleon contribution gives the solid line in
fig.~\ref{data.ps}.  The effect of the cuts is dramatic, resulting in
a very small enhancement.  Only if we take the extreme case of the
baryon density totally saturated by nucleons do we start to reach the
lower error bars of the data in the $M=200-400$ MeV regime as shown by
the dashed-dotted line.  The large $\pi N$ enhancement noted in the
rate calculation in fig.~\ref{dilep.ps}
is not present because the temperature dies away quickly, thereby
decreasing the nucleon density and rate dramatically. The fast
depletion of baryons in time
is also noted in realistic cascade and hydodynamical evolutions.

We can also evolve the photon rates and compare with the upper bounds
set by WA80 for S-Au\cite{WA80}.  The inclusion of nucleons put the
rate right on the edge of the upper limit for the data \cite{US2},
in the fire ball scenario. Since we have
analyzed both the dilepton and photon rates simultaneously, this
implies that more enhancement of the dilepton rate would overshoot the
photon data, a particularly important point in our analysis.

{\bf 5.}
For temperatures $T\sim m_{\pi}$ and densities $\rho\le 3\rho_0$, we may
treat a hadronic gas as dilute, and organize the various emission rates
using a density expansion. Assuming that the heavy-ion collisions at SPS
energies release enough entropy in the form of hadronic constituents, to
form a hadronic gas in the center of mass, we may assess reliably the photon 
and dilepton emission rates. To leading order in the density, the
rates are constrained by data and general principles. With nucleon densities
presently quoted for the SPS experiments, we are not able to account for the
present CERES data. Only larger nucleon densities (by about a factor of 4) 
may account for the data. We have checked that our results are not modified
by next-to-leading order corrections in the densities \cite{US2}. It is
important to note that our construction is not a model. Rather, models should
agree with our analysis to leading order in the pion and nucleon densities.
This construction can be extended to the higher mass region \cite{US3}.

\end{document}